\documentstyle[12pt]{article}

\begin{document}
\sloppy
\begin{flushright}{SIT-HEP/TM-3}
\end{flushright}
\vskip 1.5 truecm
\centerline{\large{\bf Electroweak baryogenesis mediated by}}
\centerline{\large{\bf locally supersymmetry-breaking defects}}
\vskip .75 truecm
\centerline{\bf Tomohiro Matsuda
\footnote{matsuda@sit.ac.jp}}
\vskip .4 truecm
\centerline {\it Laboratory of Physics, Saitama Institute of
 Technology,}
\centerline {\it Fusaiji, Okabe-machi, Saitama 369-0293, 
Japan}
\vskip 1. truecm
\makeatletter
\@addtoreset{equation}{section}
\def\theequation{\thesection.\arabic{equation}}
\makeatother
\vskip 1. truecm
\begin{abstract}
\hspace*{\parindent}
We consider the scenario of electroweak baryogenesis mediated by 
cosmological defect in models of supersymmetry breaking. 
When the effective electroweak breaking scale is raised
in the defect configuration, the mechanism of electroweak baryogenesis 
works at a higher energy scale.
The baryon charge produced by the mechanism is captured in the 
defect.
It is protected from sphalerons and then released when the defect 
decay.
\end{abstract}

\newpage
\section{Introduction}
\hspace*{\parindent}
Contrary to a naive cosmological expectation, evidences show
that the Universe contains an abundance of matter over antimatter.
In this paper we consider alternative mechanisms of electroweak 
baryogenesis.
Electroweak baryogenesis is commonly known as an attractive idea because
of its calculability in which testable physics, present in the standard 
model of electroweak interactions and its modest extensions,
is responsible for this fundamental cosmological event.
One may think that the previous negative results on the minimum standard
model as an indication that the baryon number asymmetry in the Universe
 was not created at the electroweak phase transition, but rather related 
to the physics of higher energy scales.
Of course one can stick to electroweak baryogenesis considering
the extensions of the particle content of the standard model to get 
stronger electroweak phase transition in the allowed parameter region.
In general scenario of electroweak baryogenesis the baryogenesis occurs 
at the phase boundary thus requires the co-existence of regions of 
large and small $<H/T>$, where $H$ denotes the Higgs field in the 
standard model.
In the regions of small $<H/T>$, sphalerons are unsuppressed
and can mediate baryon number violation, while the regions of large
$<H/T>$ are needed to store the created baryon number.
Below the critical temperature $T_{c}^{EW}$ of the electroweak
phase transition, $<H/T>$ grows till sphalerons are shut-off in the
whole Universe.
For electroweak baryogenesis to be possible, one needs some
specific regions where $<H>$ is displaced from the equilibrium value.

The idea we examine in this paper is that the same mechanism of
electroweak baryogenesis can happen along topological
defects left over from some other cosmological phase transitions that 
took place before the electroweak phase transition.
The idea of defect-mediated electroweak baryogenesis is already 
discussed by many authors\cite{trodden}.
They have considered the configurations of the Higgs field itself where
the vanishing Higgs vacuum expectation value is realized in the core
of the cosmological defects.
On the other hand, the displaced Higgs vacuum expectation value can be
obtained by indirect effects of other field configurations.
If the effective electroweak symmetry breaking scale is raised in some regions
inside the cosmological defect, sphalerons could be suppressed in such regions
while they would be effective in the bulk of space.
The motion of the defect network, in a similar way as
the motion of bubble walls in the usual strongly first order
phase transition scenario, will leave a net baryon number
behind the moving surface and then the baryon asymmetry will be kept 
in the sphaleron-suppressed regions inside the defect.
Then the  defect protects the baryon charge from sphalerons
until sphalerons become inactive, and then decays to release the baryon 
number after electroweak phase transition.
The idea of such protection of the produced baryon number is not new 
and commonly used by many authors in the topics
such as Q-balls \cite{Qball} or defect-mediated
baryogenesis\cite{Riotto,branden}.

In this paper, we will point out that this idea works in 
supersymmetric extensions of the standard model when the 
supersymmetry breaking
scale is raised inside the defects.
The defects should be formed before
electroweak phase transition and should decay after electroweak
phase transition.
We consider the defects that break supersymmetry locally 
and only at a short period of cosmological evolution of the Universe.

In ref.\cite{branden}, the mechanism of baryogenesis by the decay of 
the string defects which initially possess the baryon number is
discussed. 
The mechanism we will discuss in our paper can be one of the possible 
mechanisms of generating the initial baryon number asymmetry contained
in the defect.

\section{Gauge-mediated supersymmetry breaking in the defects}
\hspace*{\parindent}
Many kinds of mechanisms of breaking supersymmetry are discussed
by many authors.
The hidden sector model in supergravity\cite{sugra}
is perhaps one of the most popular scenarios among them.
In hidden sector models, supersymmetry is broken in the hidden sector
by some mechanisms, such as the Polonyi model\cite{polonyi}, 
dynamical breaking by gaugino condensation\cite{gaugino},
or the O'Raifeartaigh model\cite{O'Rafeartaigh}.
The effects of the supersymmetry breaking are mediated to the fields 
in the supersymmetric standard model only by the gravitational 
interactions, thus suppressed by the cut-off scale.

There is another mechanism of supersymmetry breaking in which the
effects are mediated by gauge interactions.
The gauge mediation of supersymmetry breaking has an attractive 
feature, ensuring the degeneracy of squark masses and
therefore suppresses the dangerous FCNC effects.

The main motivation for these models is to explain the origin and
the stability of the hierarchy between the fundamental 
scale and the electroweak scale.
In this sense, the quadratic divergence in the Higgs boson mass
parameter coming from a top quark radiative correction is cancelled
by the divergence coming from a scalar top.
Considering these cancellations and including supersymmetry breaking at 
scale $m_{SUSY}$, the resulting divergence becomes logarithmic.
The Higgs mass is reliably computed in the effective theory, and is not
dominated by unknown physics at the cutoff.
To be more precise, one can say that the stability of the hierarchy
is due to the existence of supersymmetry at higher energy scale, 
while the hierarchy is
produced by the dynamical breaking of supersymmetry or  large 
suppression factor from the fundamental scale.
One can also find some alternatives of these mechanisms of
supersymmetry breaking in the light
of M theory, large extra dimensions and brane worlds\cite{susyextra}.

Among these mechanisms of supersymmetry breaking, we first focus our 
attention to the gauge mediated models.
In this section we explore the possibility of obtaining the
baryon asymmetry of the Universe by using toy models
of gauge mediated supersymmetry breaking.

\vspace{0.3cm}
\underline{ Toy model 1}

Here we would like to consider the simplest version of the gauge mediated 
models of supersymmetry breaking which does {\it not} break supersymmetry
in the true vacuum but breaks supersymmetry locally in the defect
configuration.
The messenger sector can be described by the superpotential
\begin{equation}
\label{simplest}
W_{M}=\frac{1}{3}\lambda S^{3}-\kappa_{s}S\Lambda^{2}+
\kappa_{q}Sq\overline{q}+\kappa_{l}Sl\overline{l},
\end{equation}
where $S$ is a singlet superfield.
The superfield $q$ transforms as a (3,1,$\frac{1}{3}$) under the standard
model, while $l$ transforms as (1,2,$-\frac{1}{2}$).
Then the minima of this potential are at
\begin{eqnarray}
\label{first}
<S>&=&\pm \sqrt{\frac{\kappa_{s}}{\lambda}}\Lambda\nonumber\\
<\overline{q}q>&=&<\overline{l}l>=0
\end{eqnarray}
and 
\begin{eqnarray}
\label{second}
<S>&=&0\nonumber\\
<\overline{q}q>&\sim& <\overline{l}l>\sim \Lambda^{2}.
\end{eqnarray}
We set the scale $\Lambda$ at the intermediate scale
$\Lambda\sim 10^{6}$GeV.
$\kappa_{s},\kappa_{q}$ and $\kappa_{l}$ are assumed to be
 O(1).
$\lambda$ is not assumed to be O(1), since
the width of the domain wall is determined by $\lambda$.
Two examples are considered, $\lambda\sim 10^{-1}$ (Thin wall) and 
$\lambda\sim 10^{-6}$ (Fat wall).

At the boundary of $<\overline{q}q>=<\overline{l}l>= 0$, 
it is easy to see that the auxiliary component of the field $S$ ($F_{S}$)
can acquire vacuum expectation value of order 
$<F_{S}>\simeq \Lambda^{2}$ in the wall. 
This effect feeds down to the MSSM sector through loop corrections.
The soft terms calculated this way depend on the parameter
$<F_{S}/S>$ in the messenger sector.
In this model we assume that the effective electroweak scale in this
region becomes as large as $10^{4}$GeV. 

If the second vacuum (\ref{second}) is lifted by a small soft mass 
of $q$ and $l$, or
by contributions suppressed by the cut-off scale,
the degeneracy is broken and the domain of the local minimum
(\ref{second}) shrinks. 
It occurs when the pressure $\epsilon$  dominates the energy
density and also becomes larger than the force of the wall
tension\footnote{Here $\epsilon$ denotes
the energy difference between two minima, (\ref{first})
and(\ref{second}).}.
Denoting the induced soft mass as $m_{0}$, the temperature when the
domain shrinks is $T_{s}\simeq \sqrt{m_{0}\Lambda}$.
For $m_{0}>10^{2}$GeV, the domain shrinks before baryogenesis
starts, and the wall appears as the composite state during
baryogenesis.
The composite domain wall is constituted by two parts,
the outer region $|S|>|F_{S}|^{1/2}$ where the messenger matter fields
$q$ and $l$ acquires large mass and stay at the origin,
and the inner region $|S|<\Lambda$ where the messenger matter fields 
$q$ and $l$ can develop non-zero vacuum expectation values.
The outer side of the wall behaves almost the same as the conventional 
gauge mediation sector
of supersymmetry breaking and works as the phase boundary 
in the conventional defect-mediated electroweak baryogenesis,
producing the required gap. 
In the inner region, which we denote ``core'' of the defect,
the messenger matter fields $q$ and $l$ can develop non-zero vacuum 
expectation values and protects the incoming baryon asymmetry produced at the
outer side of the wall.

When $m_{0}$ is smaller than the electroweak scale ($m_{0}<10^{2}$GeV), 
the false vacuum $<S>=0$ appears during
the baryogenesis, and then it shrinks at $T=T_{s}$.
This period corresponds to the limit when the width of the ``core''
becomes infinite.
In the second vacuum configuration (\ref{second}), where
the gauge symmetry is already broken by non-zero expectation values of
$q$ and $l$,
the induced soft mass $m_{0}$ must be small if the supersymmetry
breaking in the bulk of space is induced by 
an another sector of gauge-mediated supersymmetry breaking.

Here we should note that the Higgs field is not the sole candidate of
the sphaleron-suppressing field that condensates inside the 
baryon-protecting defect.
It is easy to see that any field that carries the $SU(2)_{L}$ quantum numbers 
contributes to the sphaleron energy, so that the baryon number breaking 
sphaleron interactions can be suppressed by the field condensates of
other fields in the core\cite{branden}.
We should also note that if a condensate is carrying the baryon number,
then there will be massless excitations of the Goldstone
boson as well.
What we consider is the situation that the baryon number is
spontaneously broken and
the baryonic charge is stored inside the defect.
The situation is very similar to the well-known idea of B-balls or
the baryogenesis discuused in ref.\cite{branden}.

What we will concern is the situation when the effective scale
of the soft supersymmetry breaking parameter $<\left|F_{S}/S\right|>$ 
is raised
in the surface region of the defect, but the particle spectrum is
 not affected in the bulk of space.
This can be realized in a simple way if the dynamical supersymmetry
breaking sector or the messenger sector develops a cosmological defect.
In the simplest case (\ref{simplest}) the excessive breaking of 
supersymmetry is realized 
in the (composite) cosmological domain wall which interpolates two minima
$<S>=\pm \sqrt{\frac{\kappa_{s}}{\lambda}}\Lambda$.
The supersymmetry breaking in the defect sector
can vanish in the true vacuum, but should become large in the
defect to realize the co-existence of the regions of large and small
$<H/T>$.

Since the defect sector is not necessarily required to 
be responsible for the soft terms in the MSSM in the true vacuum, 
there are no complexities related to the dynamical breaking
of supersymmetry at the global minimum,
the constraint on the CP breaking parameter, etc.
\footnote{We will  comment on these issues in the appendix.}
In general, the electroweak scale is
intimately related to the soft breaking parameters
which can be raised in the ``local'' region in the defect.
\footnote{Of course, this naive expectation is not always correct.
We will also comment on this relation in the appendix,
focusing on the $\mu$ problem.}
At the temperature $T_{c}>T>T_{EW}$, baryon asymmetry produced in front
of the defect can be trapped in the defect.
Defects are able to trap the baryon from the time of the electroweak
symmetry breaking phase transition in the defect ($T=T_{c}\simeq
10^{4}$GeV) till the Universe cools down to $T=T_{EW}$.
Then the defects release the baryon number and finally
disappear at $T=T_{d}$.

Here the mechanism of baryon asymmetry generation itself is similar to 
the conventional defect-mediated electroweak baryogenesis.
Historically, the ways in which baryons may be produced when a phase 
boundary sweeps through space have been separated into 
two categories.
One is called ``local baryogenesis'' in which baryons are produced 
when the baryon number violating processes (sphaleron interactions) 
and the CP violating processes induced by the wall occur together
near the bubble walls, and the other is called ``nonlocal baryogenesis''
in which particles undergo CP violating interactions with the bubble
wall and then becomes the flux of an asymmetry which carries a quantum 
number other than the baryon number
into the unbroken phase region away from the wall.
In the latter case baryons are then produced as baryon number violating 
processes convert them into the asymmetry
 in the baryon number.
In general, both of these two ways of baryogenesis will occur together
and the baryon number asymmetry of the Universe will be expressed by 
the sum of that generated by the two coexisting processes.
When the speed of the phase boundary is greater than the sound 
speed in the plasma, local baryogenesis will dominate.
Otherwise, nonlocal baryogenesis is usually more efficient.

Let us consider the simplest case of conventional nonlocal baryogenesis,
and then examine the electroweak baryogenesis induced by the simplest
supersymmetry breaking defect.
When the thin boundary limit is considered, the final baryon to entropy 
ratio of the conventional electroweak baryogenesis becomes\cite{Riotto}
\begin{equation}
\label{nb}
 \frac{n_{B}}{s}\sim 0.2 \alpha^{2}_{W}(g^{*})^{-1}\kappa
 \Delta \theta_{CP}
\frac{1}{v_{w}} \left(\frac{m_{l}}{T}\right)^{2}
\frac{m_{h}}{T}
\frac{\xi^{L}}{D_{L}}
\end{equation}
where $D_{L}$ is the diffusion constant for leptons, and 
$\xi^{L}$ is the persistence length of the current in front of the
bubble wall.
Here $m_{l}$ and $m_{h}$ denote the lepton and Higgs masses.
When the background field configuration is steep, at the temperature much
below the phase transition in the defect at $T=T_{c}$, the effect of
CP violation is suppressed exponentially since
the typical energy of the charge carrier is lower than
the potential barrier.
In this sense the formula (\ref{nb}) can be applied in the case that the 
energy of the leptons is comparable to the Higgs VEV inside the 
phase boundary. 
In our simplest case (\ref{simplest}), the width of the defect depends on the
parameter $\lambda$.
When $\lambda\simeq$O(0.1), the width of the domain
wall is simply given by $\Delta\sim\Lambda^{-1}$, and the background
defect configuration is steep.
In a conventional scenario of electroweak baryogenesis, the flux is 
injected by the phase boundary into the unbroken phase and it is
converted
into the  baryon asymmetry in the unbroken phase near the phase boundary.
Then the produced baryons are trapped in the broken phase.
In this respect, the mechanism of baryogenesis in our model is similar 
to the conventional mechanism of electroweak baryogenesis.

The wall which interpolates between (2.2) and (2.3) can be divided into 
two parts.
In the outer half of the wall,
supersymmetry breaking parameter is locally raised so that the effective 
scale of the electroweak symmetry breaking increases simultaneously.
Considering the conventional calculation of the injected 
flux\cite{add}, one can confirm that the injected flux from the 
outer half of the domain wall into the sphaleron-activated region 
induces the baryon asymmetry which is
very similar to the conventional electroweak baryogenesis (\ref{nb}).
One may worry that the injected flux from the inner half may
cancel the one from the outer half.
Of course, it is hard to believe that they cancel exactly
even if the alternating signs of the same magnitude of injected fluxes 
are expected by the naive order estimation.
Moreover, the contribution to the injected flux from the inner half is
strongly model dependent.
It depends not only on the defect sector, but also on what one chooses
for the mechanism of supersymmetry breaking.
The important point is:
1) When the magnitude of the injected flux from the outer half is
larger than the one from the inner half, eq.(\ref{nb}) with larger
mass scales dominates.
2) When the magnitude of the injected flux from the inner half is
larger than the one from the outer half, it induces larger baryon
asymmetry. In this case, one can arrange the parameters or add new
fields in the defect sector to obtain the desirable result. 
These models will be interesting, but should be discussed in 
another place since these issues does not match our motivations of 
this paper. 
Here we do not discuss the case 2) any longer.

On the other hand, when $\lambda$ is quite small($\sim 10^{-6}$), 
the background field 
configuration can be fat enough so that the effective soft 
supersymmetry breaking mass is well approximated
by a constant around the regions of electroweak baryogenesis, then
the profile of the phase boundary is just the same
as the one in the conventional electroweak phase transition.
(Of course the effective mass scales are higher than the conventional
one.)
 The condition of such a fat background is schematically
given by the linear approximation as
\begin{equation}
m_{h}^{MAX}\times\left(\frac{T^{-1}}{\Delta}\right)\ll m_{h}^{PB},
\end{equation}
 where $m_{h}^{MAX}, \Delta$ and $m_{h}^{PB}$
denote the maximum value of $m_{h}$ inside the defect, the 
width of the wall and the Higgs mass at the phase boundary. 
Let us consider the case where the defect has the thin phase 
boundary and the fat background,
and that most of the baryon number is caught in the defect.\footnote{
Although the background changes gradually in fat defects,
the electroweak phase transition occurs at the critical point
which moves as the temperature changes.
The phase boundary can be much thinner than the background defect.}
Then we  should  integrate $n_{B}$ during the period of
$T_{c}>T>T_{end}$,
where $T_{end}$ denotes the temperature when the fat background
approximation breaks down and the suppression of the CP violation starts.
One should also consider the effective volume that the defects sweep,
whose suppression is negligible for the domain walls but can be important
to strings.

When the background wall configuration is fat, the 
baryogenesis lasts long thus the baryon number asymmetry is enhanced.
For fat walls, the energy gap which appears at the phase boundary
is determined so that the velocity of the phase boundary
equals with the speed of the background wall. 
It is easy to see that $<H>/T$ as well as the energy gap increases as 
the phase boundary moves inside, while it decreases in the
outer region.
In this respect, the critical point where the phase boundary appears
should depend on the velocity of the background domain wall.
Here we postpone the analysis on peculiar situations
($v_{w}=1$ or $v_{w}\ll 1$) but consider the
case when the velocity of the wall is close to the velocity of the
conventional electroweak phase transition.
Then $\frac{m_{l}}{T}$, $\frac{m_{h}}{T}$ and
$\frac{\xi^{L}}{D_{L}}$ in eq.(\ref{nb}) are expected to be nearly 
the same as the conventional electroweak baryogenesis.

Taking these into account, we conclude that the baryogenesis mediated by the
locally supersymmetry-breaking defects is a promising candidate
of the BAU.
The mechanism of baryogenesis itself is the same as that was used in the
conventional electroweak baryogenesis.
The novel issue in this attempt is the origin of the defects
that enables the electroweak baryogenesis at earlier (and longer)
period of the Universe when the wall is fat enough.
The electroweak phase transition itself is not required to be 
first order, which is the same characteristic of
 the conventional defect-mediated electroweak baryogenesis.

\vspace{0.3cm}
\underline{Toy model 2}

Let us consider another example of the gauge mediated model
of locally broken supersymmetry.
Here we consider the dynamical supersymmetry breaking in the
vector-like gauge theories\cite{vector}.
Denoting the singlet in the supersymmetry breaking sector by $Z$,
the low-energy effective superpotential is given by
\begin{equation}
W_{eff}=\lambda_{Z}\Lambda^{2}Z, 
\end{equation}
where $\Lambda$ is the dynamically generated scale in the supersymmetry
breaking sector.
The effective K\"ahler potential is expected to take a form
\begin{equation}
K=|Z|^{2}-\frac{\eta}{4\Lambda^{2}}|\lambda_{Z}Z|^{4}+...,
\end{equation}
where $\eta$ is a real constant of order one.
Then the effective potential is given by
\begin{equation}
V_{Z}\simeq|\lambda_{Z}|^{2}\Lambda^{4}\left(1+\frac{\eta}{\Lambda^{2}}
|\lambda_{Z}|^{4}|Z|^{2}\right).
\end{equation}
If $\eta<0$, non-zero vacuum expectation value of the singlet is
expected.
The F-component of the singlet is nonvanishing, and it is expected to be
$<F_{Z}>\simeq\lambda_{Z}\Lambda^{2}$.
The width of the defect configuration is $\Delta\simeq m_{Z}^{-1}
\simeq \left(\eta^{1/2}\lambda_{Z}^{3}\Lambda\right)^{-1}\sim (10
GeV)^{-1}$ for $F_{Z}^{1/2}=10^{6}$GeV and $\lambda_{Z}=10^{-2}$.
In this model, the fat defect can appear in a natural parameter
region.

First we consider the case when a singlet $Z$ couples directly to 
the messenger matter fields and a cosmological defect is formed
for the singlet $Z$, which is charged under global $U(1)_{R}$.
The spontaneous breakdown of the $U(1)_{R}$ symmetry in the earlier
period of the Universe produces the global string network.
The $U(1)_{R}$ symmetry is, however, must be an approximate symmetry
since it must be broken at least by an explicit
breaking constant term in the superpotential in order to
set the cosmological constant to zero.
Such an explicit breaking term induces the domain wall configuration
bounded by the string, which decays soon after it is formed.
The energy difference that lifts the degeneracy in the $U(1)_{R}$
rotation is about $\epsilon_{R}\sim
|W_{eff}|/M_{p}^{2}$ \cite{matsuda}.
Then the $U(1)_{R}$ wall-string network decays when
$T=\epsilon_{R}^{1/4}\sim 1$GeV.
The inner structure of the defect is almost the same as the domain wall
in the toy model 1.

One can also consider another example that the defect is a local
string and the configuration in the core breaks color symmetry developing
the squark vacuum expectation value.
This assumption is natural, since the (unstable) color breaking minimum
is a natural feature of the supersymmetry breaking.
Of course, one can introduce an additional defect sector which induces 
the required symmetry breaking, as we have discussed above.
In this case, the baryon number is assumed to be broken spontaneously 
inside the string, and
the baryonic charge may be stored in the core. 
Denoting the squarks as $\tilde{q}$, they carry a $U(1)$ baryonic global
charge which is derived from the conserved current
\begin{equation}
J^\mu_B=\frac{i}{2}\sum_q  q_B^q
\left(\widetilde{q}^\dagger\partial^\mu\widetilde{q}-\widetilde{q}
\partial^\mu\widetilde{q}^\dagger\right),
\end{equation}
where $q_B^q$ is the baryonic charge associated with any field
$\widetilde{q}$. 
Assuming the cylindrical symmetry, the baryonic charge per 
unit length ($Q_B$) along the $z$-axis will be given by the integration
$Q_B=\int d\theta dr\:r j_B(\theta,r)$.
This type of string is expected to generate
the suitable baryon number asymmetry of the Universe, 
if some conditions are satisfied.
In the light of ref.\cite{branden}, our mechanism of electroweak 
baryogenesis works to seed the initial baryon number 
confined in the string defects.

By interpolating two degenerated vacua in separate regions of
space, one obtains a domain wall.
If we have three or more discrete vacua in separate regions of space,
segments of domain walls can meet at a one-dimensional
``junction''.
These junctions can have the structure which is very similar to
the strings.
Although the evolution of the junctions is different from
the strings and probably much more complicated to be analyzed,
it seems possible to construct the model to produce
the baryon asymmetry in the Universe.

One can also consider an alternative of the model in which 
the mass of the messenger matter field is produced by the 
expectation values of other fields instead of $<Z>$\cite{direct}.
Then our mechanism works when the cosmological defect
is formed by the fields that generate the mass terms for messenger
matter fields $q$ and $l$.

We also note that the baryons produced by other mechanisms 
before the electroweak phase transition can survive
the wash-out if they are trapped in the supersymmetry breaking
defects that we have discussed in this paper.
This may also open another possibility for other baryogenesis.

\section{Conclusions and Discussions}
\hspace*{\parindent}
In this paper we examined new possibilities for 
electroweak baryogenesis mediated by cosmological defects.

We analyzed the supersymmetric theories in which
the hierarchy is produced by the soft breaking of supersymmetry.
Although the magnitude of the baryon asymmetry 
depends on the profiles of the defects, 
the idea is general and can be applied to many
 models of supersymmetry.
We also note that the baryons produced by other mechanisms 
before the electroweak phase transition can survive
the wash-out if they are trapped in the supersymmetry breaking
defects.
This may open another possibility for other baryogenesis.

\section{Acknowledgment}
We wish to thank K.Shima for encouragement,  and people in
Tokyo University for kind hospitality.

\appendix
\section{Constraints on cosmological domain walls}
\hspace*{\parindent}
It is well known that when the Universe undergoes a phase
transition that is associated with the spontaneous symmetry breaking of
discrete symmetries, domain walls will inevitably form.
In most cases the domain walls must be removed since they are dangerous 
for the standard evolution of the universe.
In this Appendix we make a short review to show how to estimate the constraint
to safely remove the dangerous cosmological walls.
The crudest estimate we can make will be to insist that the walls are removed 
before they dominate over the radiation energy density in the Universe.
When the explicit breaking of the discrete symmetry is expected because
of the gravitational interactions, the symmetry must be an approximate
symmetry.
Then the degeneracy of the vacua is lost and the energy difference 
$\epsilon\ne0$ appears.
When $\epsilon$ dominates the energy density of the false
vacuum, regions of the higher density false vacuum tend to shrink.
The corresponding force per unit area of the wall is $\sim \epsilon$.
The energy difference $\epsilon$ becomes dynamically 
important when this force becomes comparable to the force of 
the tension $f\sim \sigma/R_{w}$,
where $\sigma$ is the surface energy density of the wall
and $R_{w}$ denotes the typical scale of the wall distance.
For walls to disappear safely, this has to happen before the
walls dominate the Universe.
On the other hand, the domain wall network is not a static 
system.
In general, the initial shape of the walls right after the
phase transition is determined by the random variation
of the scalar VEV.
One may expect the walls just after they are formed 
to be very irregular, random
surfaces with a typical curvature radius, which
is determined by the correlation length of the
scalar field.
To characterize the system of domain walls,
simulations\cite{simulation} are commonly used.
According to the simulations, the system will be dominated by 
one large (infinite size)
wall network  and some finite closed walls (cells) just after the phase
transition.
The isolated closed walls smaller than the horizon
will shrink and disappear soon after they are formed.
Since the walls smaller than the horizon size 
will efficiently disappear so that only walls
at the horizon size will remain,
their typical curvature scale will be the horizon 
size, $R\sim t\sim M_{p}/g_{*}^{\frac{1}{2}}T^{2}$.
Then the energy density of the wall $\rho_{w}$ is about
\begin{equation}
\rho_{w}\sim \frac{\sigma}{R},
\end{equation}
and the radiation energy density $\rho_{r}$ is 
\begin{equation}
\rho_{r}\sim g_{*}T^{4},
\end{equation}
one can see that the wall domination starts below a temperature $T_{w}$
\begin{equation}
T_{w}\sim \left(\frac{\sigma}{g_{*}^{1/2}M_{p}}
\right)^{\frac{1}{2}}.
\end{equation}
To prevent the wall domination, one requires the
pressure $\epsilon$ to have become dominant before this epoch.
This requires the constraint,
\begin{equation}
\label{criterion}
\epsilon>\frac{\sigma}{R_{wd}}\sim
\frac{\sigma^{2}}{M^{2}_{p}}.
\end{equation}
Here $R_{wd}$ denotes the horizon size at the wall domination.
A pressure of this magnitude is expected to be produced by
higher dimensional operators which explicitly break
the discrete symmetry.
The requirement is satisfied in general models of supersymmetry
if the symmetry is a discrete R-symmetry $Z^{R}_{n}$\cite{matsuda}.

In our model, the requirement that the walls decay after 
electroweak symmetry breaking also
imposes the upper bound on $\sigma$ as $\sigma<(10^{8}GeV)^{3}$,
which excludes the hidden ($M_{p}$ suppressed) sector for
the defect.

Here we should note about the lower bound that comes from the
nucleosynthesis.
The criterion (\ref{criterion})
seems appropriate, if the scale of the wall is higher 
than $(10^{5}GeV)^{3}$.
For the walls below this scale ($\sigma\le(10^{5}GeV)^{3}$),
 there should be  further constraints coming from primordial 
nucleosynthesis.
Since the time associated with the collapsing temperature
 $T_{w}$
is $t_{w}\sim M_{p}^{2}/g_{*}^{\frac{1}{2}}\sigma
\sim 10^{8}\left(\frac{(10^{2}GeV)^{3}}{\sigma}\right)$sec,
the walls $\sigma\le(10^{5}GeV)^{3}$ will decay after 
nucleosynthesis\cite{Abel} and violate the phenomenological bounds
for nucleosynthesis.
If the walls are not hidden and can decay into the standard
model particles, the entropy produced when walls collapse
will violate the phenomenological bounds for nucleosynthesis.
On the other hand, 
the succeeding story should strongly depend on the 
details of the hidden components and their interactions if the walls are
 soft domain walls\cite{soft_wall}.
They can decay late to 
contribute to the large scale structure formation.

Of course, the condition for the cosmological domain wall not to dominate
the Universe (\ref{criterion}) 
should also be modified if the wall velocity is lower than
the speed of the light and then the Universe contains walls more than one. 
This implies that the condition to evade the wall domination becomes 
$\epsilon>(\sigma^{2}/M_{p}^{2})\times x$,
where the constant $x$ is determined by $R_{w}$ as 
$x\simeq M_{p}/(R_{w}T^{2})$.
For the walls with lower velocity, the bound for $\epsilon$ 
is inevitably raised since such walls will dominate earlier.

\section{$\mu$ term and CP violation}
\hspace*{\parindent}
In order to make our discussions simple and generic, we made a naive 
assumption 
that the relative relations between mass parameters that appear in the 
conventional scenario for electroweak baryogenesis are not drastically
changed at the surface of the defect so that one can use the conventional 
mechanism of electroweak 
baryogenesis. 
When the mechanism of generating the $\mu$ and $B_{\mu}$-term 
in the defect is completely
different from the one for the supersymmetry breaking soft masses,
the $\mu$ and $B_{\mu}$-term are in general not altered and remain the
same in the locally supersymmetry-breaking defects.
Then the effective soft supersymmetry breaking mass becomes 
$10\sim10^{2}$ times
larger than the $\mu$ and $B_{\mu}$-term in the effective theory at the
defect surfaces.
One may also consider other alternatives in which the structure of the
effective low energy theory becomes completely different in the defect.
These models will be interesting, but should be discussed in another place
since these issues does not match our motivations of this paper.

Of course, one knows in some models the
$\mu$ or $B_{\mu}$-term can be related to the supersymmetry breaking
parameters $F_{X}/X$ \cite{mu_dvali,moroi}.
The most interesting case is that the $\mu$-term originates from
the interaction with the supersymmetry breaking sector,
while the $B_{\mu}$ is suppressed so that the SUSY CP problem
is solved in the bulk of space.
In such models for $\mu$ and $B_{\mu}$-term generation, almost
all the input mass scales are determined by the supersymmetry
breaking parameter $F_{X}/X$ and the mechanism of the electroweak 
baryogenesis  at the defect
surface looks precisely the same as the one for the conventional MSSM.

One can construct models in which the generating mechanism of
$B_{\mu}$ in the defect is different from the one in the bulk of 
space, so that the effective theory
at the defect surface can develop large $B_{\mu}$-parameter. 
In this case, $B_{\mu}^{eff}$ locally induces large CP parameter and
the electroweak baryogenesis is enhanced.\footnote{
In the gauge mediated model, the phases in the gaugino masses $m_{G}$,
$\mu$ and $B_{\mu}$ parameters are physical, and in general,
they can be large enough to conflict with experimental
constraints.
There are many ways to restrict these CP violating phases.
For example, the SUSY CP problem can be solved if
$B_{\mu}$ vanishes at the messenger scale, which is further
discussed in ref.\cite{bmu}.}

In our model for electroweak baryogenesis, one can expect
another contribution to these CP violating phases
from locally supersymmetry breaking defects, since
the origin of the supersymmetry breaking mass
and $\mu$ and $B_{\mu}$ parameters
are completely or partially different from the true vacuum,
changing the combination 
$\theta_{phys}\equiv Arg(\mu B_{\mu}^{*} m_{G})=\theta_{\mu}-\theta_{B}
+\theta_{G}$.

\end{document}